\documentclass
[floatfix,superscriptaddress,secnumarabic,amssymb,amsmath,nobibnotes,aps,prd,showkeys,nofootinbib,twocolumn,notitlepage,10pt]{revtex4}%
\usepackage{setspace}
\usepackage{xcolor}
\usepackage{amsmath}
\usepackage{amsfonts}
\usepackage{amssymb}
\usepackage{verbatim}
\usepackage{graphicx,bm}
\usepackage[caption=false]{subfig}
\usepackage[colorlinks]{hyperref}
\usepackage{graphicx}%
\providecommand{\U}[1]{\protect\rule{.1in}{.1in}}

\newcommand{\be}{\begin{equation}}
\newcommand{\ee}{\end{equation}}

\newcommand{\mincir}{\raise
-3.truept\hbox{\rlap{\hbox{$\sim$}}\raise4.truept\hbox{$<$}\ }}
\newcommand{\magcir}{\raise
-3.truept\hbox{\rlap{\hbox{$\sim$}}\raise4.truept\hbox{$>$}\ }}

\hypersetup{
breaklinks=true,
pdfstartview={FitH},      colorlinks=true,           linkcolor=blue,              citecolor=red,            filecolor=magenta,          urlcolor=blue,               anchorcolor=green,          linktocpage=true
}
\ifx\pdfoutput\relax\let\pdfoutput=\undefined\fi
\newcount\msipdfoutput
\ifx\pdfoutput\undefined\else
\ifcase\pdfoutput\else
\ifx\paperwidth\undefined\else
\ifdim\paperheight=0pt\relax\else\pdfpageheight\paperheight\fi
\ifdim\paperwidth=0pt\relax\else\pdfpagewidth\paperwidth\fi
\fi\fi\fi
\begin{document}
\title{Observational constraints on Barrow holographic dark energy
coupled with a non-cold dark matter component from DESI DR2}
\author{Abdulla Al Mamon}
\email{abdulla.physics@gmail.com}
\affiliation{Department of Physics, Vivekananda Satavarshiki Mahavidyalaya (affiliated to
the Vidyasagar University), Manikpara-721513, West Bengal, India}
\author{Genly Leon}
\email{genly.leon@ucn.cl}
\affiliation{Departamento de Matem\'{a}ticas, Universidad Cat\'{o}lica del Norte, Avenida
Angamos 0610, Casilla 1280 Antofagasta, Chile}
\affiliation{Institute of Systems Science, Durban University of Technology, Durban 4000,
South Africa}
\affiliation{Centre for Space Research, North-West University, Potchefstroom 2520, South Africa}
\author{Subhajit Saha}
\email{subhajit1729@gmail.com}
\affiliation{Department of Mathematics, Panihati Mahavidyalaya, Kolkata 700110, West
Bengal, India}
\author{Andronikos Paliathanasis}
\email{anpaliat@phys.uoa.gr}
\affiliation{Institute of Systems Science, Durban University of Technology, Durban 4000,
South Africa}
\affiliation{Centre for Space Research, North-West University, Potchefstroom 2520, South Africa}
\affiliation{Centro de Investigaci\'on, Innovaci\'on y Creaci\'on (CIIC), Universidad
Cat\'olica de Temuco, Temuco, Chile}
\affiliation{Departamento de Ciencias Matem\'{a}ticas y F\'{\i}sicas, Facultad de
Ingenier\'{\i}a, Universidad Cat\'olica de Temuco, Temuco, Chile}
\affiliation{National Institute for Theoretical and Computational Sciences (NITheCS), South Africa}

\begin{abstract}
We investigate the cosmological viability of Barrow holographic dark energy in
a spatially flat Friedmann--Lema\^{\i}tre--Robertson--Walker universe in which
the dark matter component is allowed to have a non-zero pressure,
characterized by a constant equation-of-state parameter $w_{m}$. By
considering the future event horizon as the infrared cutoff, we derive the
master equation, which describes the cosmological dynamics for the background
space. We constrain the model against late-time data, combining the baryon
acoustic oscillation from DESI DR2 with three different catalogues for the
Type Ia Supernova measurements and the Cosmic Chronometers. The dark matter equation of state is constrained to $w_{m}=0.033_{-0.030}^{+0.045}$, $0.009_{-0.041}%
^{+0.047}$, and $0.033_{-0.029}^{+0.042}$, for the PantheonPlus, the Union3.0
and the DES-Dovekie supernova datasets respectively. Therefore, the pressureless limit is recovered within the $2\sigma$ regime. On the other hand, the Barrow exponent is weakly constrained, due to the $\Delta-w_{m}$ degeneracy. The dark energy equation of
state remains above the phantom divide throughout the redshift range probed.
Finally, in the comparison of the statistical parameters with that of 
$\Lambda$CDM, it follows $\Delta\mathrm{AIC}=+2.23$, $+0.75$, and $+1.38$,
$\ $while for the Bayesian evidence we find $\Delta\ln Z=-0.63$, $+0.47$, and
$-0.13$, which suggest that the datasets considered in this analysis do not have a preferred model. Finally the relation with the corresponding Tsalis holographic dark energy model is discussed. 

\end{abstract}

\maketitle

\section{Introduction}

\label{sec1}
Observations of the late-time Universe show that its energy content is
overwhelmingly dominated by two dark components: dark matter (DM) and dark
energy (DE), which together make up about 95\% of the total cosmic energy
budget~\cite{refplanck}. Whereas DM is essential for the growth and clustering
of cosmic structures~\cite{refdm01}, DE is invoked to explain the present
accelerated expansion of the Universe~\cite{acc1,acc2}. Nevertheless, the
fundamental physical nature of both components has not yet been understood at
a deeper level. Within the standard $\Lambda$CDM paradigm, DE is represented
by a cosmological constant $\Lambda$ with a fixed equation of state (EoS) $w =
-1$, while dark matter is modeled as a cold, pressureless fluid with zero EoS
$w_{\mathrm{dm}} = 0$, interacting with other components only through gravity.
Although the $\Lambda$CDM model has achieved remarkable success, it is still
faced with a number of theoretical and observational difficulties,
especially those related to the cosmological constant problem and the physical
nature of cold dark matter \cite{Perivolaropoulos1}.

Alternatively, various extensions of the standard cosmological model have been
proposed. Among them, holographic dark energy (HDE) scenarios, motivated by
ideas from quantum gravity, stand out as a promising class of alternatives. In
particular, HDE provides a different theoretical framework for explaining DE,
grounded in the holographic principle \cite{refhp1,refhp2,refhp3,refshde1}.
More recently, guided by developments in generalized statistical mechanics,
several extended HDE models have been formulated using modified entropy
formalisms. In particular, HDE models based on Barrow \cite{refbarrowen},
Tsallis \cite{refte1,refte2,reftebook}, and Viaggiu \cite{refve1,refve2}
entropies have attracted significant attention. The cosmological consequences
of these generalized HDE constructions have been thoroughly examined in the
literature \cite{refbhde,refbhde2,refbhde3,
refbhde4,refbhde5,refbhde6,refbhde7,refbhde8,refbhdedv1,refbhdedv2,refthde1,refthde2,refthde3,refthde4,refthde5,refthde6,refthde7,refvhde1,refvhde2,refvhde3}%
. For a comprehensive review of the HDE paradigm, one may refer to the extensive
review \cite{refshde2}, which trace the theoretical development of HDE from its
basis in the holographic principle through its observational constraints and
extensions, including interacting scenarios with DM. It is worth noting that a
generalized horizon entropy has recently been introduced, linking the field
equations of an FLRW universe in a general gravity theory to the thermodynamic
behavior of the dynamical apparent horizon
\cite{Nojiri1,Nojiri2,Luciano1,Luciano2,Leizerovich}. Most of the entropy
formalisms discussed in the literature, including those of Barrow and Tsallis,
turn out to be particular cases of this more general form.

Analysis of recent observations from the Dark Energy Spectroscopic Instrument (DESI), based on baryon acoustic oscillation (BAO) measurements from its second data release (DR2), have provided compelling indications in favor of dynamical DE, see for example \cite{desidr22dde}. These results suggest that
the DE EoS may evolve with cosmic time, thereby challenging the standard
$\Lambda$CDM paradigm. It is important to note that such conclusions are
typically obtained under the conventional assumption that DM behaves as a
perfectly cold fluid characterized by a vanishing EoS.

This raises a fundamental question regarding the validity of this assumption.
If DE is allowed to evolve, it is natural to explore whether DM may also
deviate from its standard cold and pressureless description. Given that both
DM and DE play essential roles in determining the expansion history of the
Universe as well as the formation and growth of large-scale structures, their
physical properties are inherently coupled at the level of cosmological
dynamics. This consideration motivates a more general framework in which both
components of the dark sector are treated as dynamical entities, potentially
offering a richer and more complete description of the Universe.

Motivated by the theoretical framework introduced in Refs.
\cite{ddmmoti1,ddmmoti2}, several studies have explored the possibility that
observational data may favor non-cold DM with a non-zero EoS parameter; see,
for instance,
Refs.~\cite{ddm1,ddm2,ddm3,ddm4,ddm5,ddm6,ddm7,ddm8,ddm9,ddm10,ddm11,ddm12,ddm13,ddm14,ddm15,ddm16,ddm17,ddmpara,refliddm}%
. It is worth highlighting that, although previous works have investigated the
possibility of a time-varying DM EoS, these studies have typically been
limited to a narrow region of parameter space, which may obscure potential
imprints of new physics \cite{ddm2,ddm3}. More recently, in~\cite{ddmpara} it
addressed this limitation by extending the parameter space of the DM EoS to
its full range, rather than simply incorporating updated observational
datasets. Their analysis, based on individual probes including the cosmic
microwave background (CMB), observations from DESI, and Pantheon+, reveal
hints of dynamical DM at an approximately $2\sigma$ confidence level.
Furthermore, when CMB data are combined with low-redshift observations, the
significance exceeds the $2\sigma$ threshold, thereby indicating a possible
scenario in which both dynamical DM and dynamical DE coexist within the dark
sector. In a separate study, in \cite{refliddm} reported that recent
observations from DESI suggest that DM may deviate from the standard cold
assumption and could possess a non-zero EoS parameter. At the same time, their
results also indicate that DE may not be consistent with a cosmological
constant, but instead may exhibit dynamical behavior.

In the present work, we investigate two different entropy-inspired HDE
frameworks, namely the Barrow and Tsallis HDE models, using DESI DR2 data,
together with dynamical dark sector scenarios in which the dark matter has a
nonzero pressure component. A detailed discussion of the Barrow and Tsallis
HDE models is presented in the Section \ref{sec2}, where we present the basic
equations which describe the background dynamics. We focus on testing the
observational viability of these models and constraining their parameters
using current late-time cosmological data. In particular we employ a
combination of Supernova data, of the direct measurements of the Hubble
parameter from the Cosmic Chronometers and the BAO measurements.\ The Bayesian
analysis in presented in Section \ref{sec4}. Finally in Section \ref{sec5} we
draw our conclusions.


\section{Theoretical framework and model formulation}

\label{sec2}

\subsection{Basic cosmological equations}

\label{sec21}
We begin with a spatially flat Friedmann-Lema\^{\i}tre-Robertson-Walker metric
(FLRW) spacetime, whose line element is given by
\begin{equation}
ds^{2}=-dt^{2}+a^{2}(t)\,\delta_{ij}\,dx^{i}dx^{j},
\end{equation}
where $a(t)$ is the cosmic scale factor at time $t$, normalized such that
$a(t_{0})=1$, with $t_{0}$ denoting the present time.

The gravitational dynamics are assumed to follow Einstein's General
Relativity, with matter fields minimally coupled to gravity. Furthermore, we
assume that the various components do not interact through non-gravitational
processes, so that each fluid evolves independently. In this setup, the
dynamics of cosmic expansion is governed by the Friedmann equations \cite{de1},%

\begin{align}
\label{eq-fe1}H^{2}  &  =\frac{1}{3M_{p}^{2}}\sum_{i} \rho_{i}, \qquad
i=(m,b,d),\\
\dot{H}  &  =-\frac{1}{2M_{p}^{2}}\sum_{i} (\rho_{i} + p_{i}). \label{eq-fe2}%
\end{align}

In these expressions, the Hubble parameter is given by $H=\frac{\dot{a}}{a}$,
an overdot denotes differentiation with respect to cosmic time $t$ and
$M_{p}\equiv(8\pi G)^{-\frac{1}{2}}$ is the reduced Planck mass. The
quantities $\rho_{i}$ and $p_{i}$ correspond to the energy density and
pressure of the $i$-th component, namely non-cold dark matter $(m)$,
pressureless baryons $(b)$, and HDE $(d)$. Accordingly, the total energy
density of the Universe can be written as the sum of its individual
components:
\begin{equation}
\rho= \rho_{m}+\rho_{b}+\rho_{d},
\end{equation}
while the total pressure is given by
\begin{equation}
p=p_{m}+p_{b}+p_{d}.
\end{equation}
We further assume that there are no non-gravitational interactions among these
components; therefore, each fluid evolves independently and satisfies its own
conservation equation. Consequently, the continuity equation for each
component can be written as
\begin{equation}
\label{eq-cons}\dot{\rho}_{i} + 3H(\rho_{i} + p_{i}) = 0,
\end{equation}
Each component is also characterized by an EoS parameter defined as
\begin{equation}
w_{i} = \frac{p_{i}}{\rho_{i}}.
\end{equation}
For the matter component, we take the EoS $p_{m}=w_{m}\rho_{m}$ with
$w_{m}\neq0$, whereas baryons are assumed to be pressureless $p_{b}=0$. Under
these conditions, Eq. (\ref{eq-cons}) gives the standard evolution laws
\begin{align}
\label{eq-rhoma1}\rho_{m}  &  = \rho_{m0}\, a^{-3(1+w_{m})},\\
\rho_{b}  &  = \rho_{b0}\, a^{-3}. \label{eq-rhomb1}%
\end{align}
From this point onward in the text, the subscript ``$0$'' represents the
present-day value of a given quantity. In addition, the DE component is
governed by the continuity equation
\begin{equation}
\label{eq-wdcons}\dot{\rho}_{d} + 3H(1+w_{d})\rho_{d} = 0 .
\end{equation}
Rewriting this relation allows the DE EoS parameter to be expressed as
\begin{equation}
\label{eq-wdgen}w_{d} = -1 - \frac{\dot{\rho}_{d}}{3H\rho_{d}}.
\end{equation}
This provides a direct relationship between the evolution of the DE density
and its effective EoS.

Now, the Raychaudhuri equation (\ref{eq-fe2}), which governs the evolution of
the Hubble parameter, can be rewritten as follows:
\begin{equation}
\label{eq-fe2rho}\dot H=-\frac{1}{2M_{p}^{2}}\left[  (1+w_{m})\rho_{m}%
+\rho_{b}+(1+w_{d})\rho_{d}\right]  .
\end{equation}
Using the Friedmann equations together with the conservation relations
introduced above, it is convenient to recast the cosmic dynamics in terms of
dimensionless density parameters. The dimensionless density parameters are
defined as
\begin{equation}
\label{eq-Odefinition}\Omega_{i}=\frac{\rho_{i}}{3M_{p}^{2}H^{2}}, \qquad
i=(m,b,d),
\end{equation}
and they satisfy the normalization condition;
\begin{equation}
\label{eq-normalization condition}\Omega_{m}+\Omega_{b}+\Omega_{d}=1 .
\end{equation}
From Eqs.~\eqref{eq-rhoma1},~\eqref{eq-rhomb1} and \eqref{eq-Odefinition}, the
matter and baryonic density parameters are expressed as
\begin{equation}
\label{eq-Omegamawm}\Omega_{m}=\frac{H^{2}_{0}\Omega_{m0}a^{-3(1+w_{m})}%
}{H^{2}}, \qquad\Omega_{b}=\frac{H^{2}_{0}\Omega_{b0}a^{-3}}{H^{2}}.
\end{equation}
Substituting Eq. (\ref{eq-Omegamawm}) into Eq.
(\ref{eq-normalization condition}) and rearranging the resulting equation
yields
\begin{equation}
\label{eq-H2gen001}H^{2}= \frac{H_{0}^{2} \left[  \Omega_{m0}a^{-3(1+w_{m})}
+\Omega_{b0}a^{-3} \right]  } {1-\Omega_{d}},
\end{equation}
or equivalently,
\begin{equation}
\label{eq-Hgen001}\frac{1}{Ha}=\frac{a^{\frac{1}{2}}\sqrt{(1-\Omega_{d})}%
}{H_{0}\sqrt{(\Omega_{m0}a^{-3w_{m}} + \Omega_{b0})}}.
\end{equation}
Consequently, the baryonic density parameter can be rewritten in the following
form:
\begin{align}
\label{eq-OmegabaryonGen}\Omega_{b}  &  = \frac{\Omega_{b0}(1-\Omega_{d}%
)}{\Omega_{m0}e^{-3w_{m}x}+\Omega_{b0}}.
\end{align}
with $x=\ln a$, or equivalently, $a=e^{x}$.

Expressing the energy densities as $\rho_{i}=3M_{p}^{2}H^{2}\Omega_{i}$, the
evolution of the Hubble parameter can then be written as%

\begin{equation}
\label{eq-HdotH2}\frac{H^{\prime}}{H}=\frac{\dot{H}}{H^{2}}=-\frac{3}{2}
\left[  1 + w_{m}(1-\Omega_{b}-\Omega_{d}) + w_{d}\Omega_{d} \right]  .
\end{equation}
where a prime denotes differentiation with respect to the natural logarithm of
the scale factor. The above equation describes the evolution of the Hubble
expansion rate in a universe consisting of baryonic matter, a matter with
non-zero pressure component, and DE. In this case, matter behaves like
pressureless dust ($w_{m} = 0$), the evolution of the Hubble parameter is
determined entirely by the DE EoS parameter, and its fractional energy
density.

In the subsequent subsections, we provide a concise overview of two different
HDE models, namely Barrow and Tsallis HDE, considered within the cosmological
framework introduced above.


\subsection{Barrow holographic dark energy with matter}

\label{sec24}
This section provides a brief overview of the Barrow holographic dark energy
(BHDE) framework \cite{refbhde}. According to the Barrow formulation,
departures from the standard Bekenstein-Hawking entropy are understood as a
consequence of quantum gravitational corrections, which can give rise to a
fractal-type structure on black holes as well as cosmological horizons. In
this scenario, quantum fluctuations are effectively characterized by a
dimensionless parameter $\Delta$ ($0\leq\Delta\leq1$), which quantifies the
degree of spacetime irregularity on the horizon. Accordingly, the entropy-area
relation is modified and can be written as \cite{refbarrowen}
\begin{equation}
S_{\Delta}=\left(  \frac{A}{A_{0}}\right)  ^{1+\frac{\Delta}{2}}%
.\label{eq-BhdeEntropy}%
\end{equation}
In this expression, the quantum deformation is encoded in the exponent
$\Delta$, which lies within the interval $\Delta\in\lbrack0,1]$. In
particular, $\Delta=1$ represents the case of maximal deformation, while
$\Delta=0$ corresponds to the simplest horizon configuration, in which the
standard Bekenstein entropy is recovered \cite{refbhe1,refbhe2}. The standard
HDE model \cite{refshde1,refshde2} is based on the entropy bound $\rho
_{d}L^{4}\leq S$, where $\rho_{d}$ denotes the energy density associated with
HDE and $L$ represents the infrared (IR) cutoff or the horizon scale. Using
the usual area scaling $S\propto A\propto L^{2}$~\cite{refshde1,refshde2},
replacing the entropy with the Barrow form in Eq.~(\ref{eq-BhdeEntropy}),
leads to the expression \cite{refbhde}
\begin{equation}
\rho_{d}=CL^{\Delta-2},\label{eq-rhobhde}%
\end{equation}
where $C$ is a constant parameter. The characteristic length scale or the IR
cutoff $L$ can be chosen in several different ways depending on the physical
setup. A commonly used and phenomenologically successful choice is the future
event horizon \cite{refshde1,refshde2}. Consequently, in this work, we
identify the IR cutoff with the future event horizon radius $R_{E}$, which is
mathematically defined as the integral
\begin{equation}
L=R_{E}=a\int_{t}^{\infty}\frac{dt^{\prime}}{a(t^{\prime})}=a\int_{a}^{\infty
}\frac{da^{\prime}}{H{a^{\prime}}^{2}}=a\int_{x}^{\infty}\frac{dx}%
{Ha}.\label{eq-feh}%
\end{equation}
Taking the time derivative of Eq. (\ref{eq-feh}) gives
\begin{equation}
\dot{L}=HL-1.\label{eq-Ldotm}%
\end{equation}
In this scenario, the dimensionless density parameter is expressed as
\begin{equation}
\Omega_{d}\equiv\frac{\rho_{d}}{3M_{p}^{2}H^{2}}=\frac{CL^{\Delta-2}}%
{3M_{p}^{2}H^{2}}.\label{eq-Odbhde001}%
\end{equation}
Consequently, the above expression can be expressed in terms of the variable
$x$ as
\begin{equation}
\int_{x}^{\infty}\frac{dx}{Ha}=\frac{1}{a}\left(  \frac{{C}}{3M_{p}^{2}%
H^{2}\Omega_{d}}\right)  ^{\frac{1}{2-\Delta}},\label{eq-integrrelation}%
\end{equation}
Now, substituting the expression for $\frac{1}{Ha}$ obtained from
Eq.~(\ref{eq-Hgen001}) into Eq.~(\ref{eq-integrrelation}), we obtain
\begin{equation}
\int_{x}^{\infty}\frac{e^{\frac{x}{2}\sqrt{1-\Omega_{d}}}}{H_{0}\sqrt
{(\Omega_{m0}e^{-3w_{m}x}+\Omega_{b0})}}dx=\frac{1}{a}\left(  \frac{{C}%
}{3M_{p}^{2}H^{2}\Omega_{d}}\right)  ^{\frac{1}{2-\Delta}}%
.\label{eq-integrrelation2}%
\end{equation}
By differentiating Eq. (\ref{eq-integrrelation2}) with respect to $x=\ln a$,
we arrive at
\begin{widetext}
\begin{equation}
\!\!\!\!\!\!\!\!\!\!\!\!\!\frac{\Omega_{d}^{\prime}}{\Omega_{d}(1-\Omega_{d}%
)}=1+\Delta+3w_{m}\frac{\Omega_{m0}e^{-3w_{m}x}}{(\Omega_{m0}e^{-3w_{m}%
x}+\Omega_{b0})}+Q_{eff}(x)(1-\Omega_{d})^{\frac{\Delta}{2(\Delta-2)}}%
\cdot(\Omega_{d})^{\frac{1}{2-\Delta}}e^{\frac{3\Delta}{2(\Delta-2)}%
x},\label{eq-OdediffeqBhde}%
\end{equation}
\end{widetext}
where 
\begin{widetext}
\begin{equation}
Q_{eff}(x)=(2-\Delta)\left(  \frac{{C}}{3M_{p}^{2}}\right)  ^{\frac
{1}{\Delta-2}}\left(  H_{0}\right)  ^{\frac{\Delta}{2-\Delta}}\left(
\Omega_{m0}e^{-3w_{m}x}+\Omega_{b0}\right)  ^{\frac{\Delta}{2(2-\Delta)}}
\end{equation}
\end{widetext} is
and the prime denotes differentiation with respect to $x$.  

The differential equation derived above describes the evolution of BHDE in a spatially flat universe containing matter with non-zero pressure. For the special case $\Delta=0$, $w_{m}=0$, $\Omega_{b0}=0$ the model reduces to the standard HDE scenario, and the corresponding evolution equation takes the form
\begin{equation}
\Omega_{d}^{\prime}\big|_{\Delta=0}=\Omega_{d}(1-\Omega_{d})\left[
1+2{\left(  \frac{3M_{p}^{2}\Omega_{d}}{C}\right)  }^{\frac{1}{2}}\right]  .
\end{equation}

In this limit, the equation admits an implicit analytical solution
\cite{refshde1,refshde2}. Furthermore, in the limit of a nonzero Barrow
exponent $\Delta$, with $w_{m}=0$ and $\Omega_{b0}=0$, our evolution equation
for $\Omega_{d}$ reduces exactly to the evolution equation obtained in the
original BHDE model, where the parameter $Q_{eff}$ becomes the constant $Q$
defined in Ref. \cite{refbhde}. However, for general values of the Barrow
exponent $\Delta$, with $w_{m}\neq0$ and $\Omega_{b0}\neq0$,
Eq.~(\ref{eq-OdediffeqBhde}) contains an explicit dependence on
$x=\mbox{ln}~a$, which prevents the derivation of a closed-form analytical
solution. Therefore, the evolution of the DE density parameter must be
investigated numerically.

Using the above relations, the EoS parameter of the BHDE model, $w_{d}$, can
be derived. By differentiating the DE density given in Eq.~(\ref{eq-rhobhde})
with respect to cosmic time and making use of Eqs.~(\ref{eq-wdcons}) and
(\ref{eq-Ldotm}), we obtain the following expression for the EoS parameter:
\begin{equation}
w_{d}=-\frac{1}{3}-\frac{\Delta}{3}+\frac{\Delta-2}{3HR_{E}}%
.\label{eq-wdbhde001}%
\end{equation}

\noindent According to Eq.~(\ref{eq-Odbhde001}), the quantity $\frac{1}%
{HR_{E}}$ can be expressed as
\begin{equation}
\frac{1}{HR_{E}}=H^{-1}\left(  \frac{3M_{p}^{2}H^{2}\Omega_{d}}{C}\right)
^{\frac{1}{2-\Delta}}.
\end{equation}
Now, substituting the expression for $H$ from Eq.~(\ref{eq-Hgen001}), we
obtain
\begin{widetext}
\begin{equation}
\frac{1}{HR_{E}}=\left(  \frac{3M_{p}^{2}}{C}\right)  ^{\frac{1}{2-\Delta}%
}H_{0}^{\frac{\Delta}{2-\Delta}}\Omega_{D}^{\frac{1}{2-\Delta}}(1-\Omega
_{D})^{\frac{\Delta}{2(\Delta-2)}}\left(  \Omega_{m0}e^{-3w_{m}x}+\Omega
_{b0}\right)  ^{\frac{\Delta}{2(2-\Delta)}}e^{-\frac{3\Delta}{2(2-\Delta)}x}.
\end{equation}
\end{widetext}
Therefore, using the above expression for $\frac{1}{HR_{E}}$, we finally
obtain
\begin{equation}
w_{d}=-\frac{1+\Delta}{3}-\frac{Q_{eff}(x)}{3}\left(  \Omega_{d}\right)
^{\frac{1}{2-\Delta}}\left(  1-\Omega_{d}\right)  ^{\frac{\Delta}{2(\Delta
-2)}}e^{\frac{3\Delta}{2(\Delta-2)}x}.
\end{equation}
Thus, once the evolution of $\Omega_{d}$ is determined by numerically solving
the corresponding differential equation, the EoS parameter $w_{d}$ can be
obtained directly. Again, in the limiting case $\Delta=0$, $w_{m}=0$, and
$\Omega_{b0}=0$, the present model recovers the standard HDE scenario
\cite{refshde1,refshde2}. In this limit, the EoS parameter $w_{d}$ becomes
\begin{equation}
w_{d}\big|_{\Delta=0}=-\frac{1}{3}-\frac{2}{3}\left(  \frac{3M_{p}^{2}%
\Omega_{d}}{C}\right)  ^{\frac{1}{2}}.
\end{equation}
Similarly, for a nonzero Barrow exponent $\Delta$, with $w_{m}=0$ and
$\Omega_{b0}=0$, the evolution equation for $w_{d}$ reduces to the
corresponding result of the original BHDE model. In this limit, the effective
parameter $Q_{\mathrm{eff}}$ becomes the constant $Q$ introduced in Ref.
\cite{refbhde}.


\subsection{Tsallis holographic dark energy model with matter}

\label{sec23}
This subsection introduces a modified DE model derived from Tsallis entropy
combined with the holographic principle, commonly referred to as Tsallis
holographic dark energy (THDE) \cite{refthde1}. The Tsallis entropy is given
by \cite{refte1,refte2,reftebook}
\begin{equation}
S_{\mu}=\left(  \frac{A}{A_{0}}\right)  ^{\mu}, \label{eq-thdeEntropy}%
\end{equation}
where $A$ denotes the usual horizon area, while $A_{0}$ corresponds to the
Planck area. The parameter $\mu$ characterizes the deviation from the standard
Bekenstein-Hawking area law and incorporates possible effects arising from the
underlying theoretical framework. The conventional Bekenstein-Hawking entropy
is recovered when $\mu=1$. Within the Tsallis framework, the parameter $\mu$
quantifies the extent of non-additivity in the system. When $\mu<1$, the
system exhibits sub-additive behavior, which means that the number of
accessible microstates grows more slowly than the horizon area, resulting in a
slower increase of entropy with system size. In contrast, $\mu>1$ corresponds
to a super-additive regime, which may be interpreted as an effective
enhancement in the number of degrees of freedom or the presence of nonlocal or
high-energy effects.

One can note that the Barrow entropy $S_{\Delta}$ in Eq.~(\ref{eq-BhdeEntropy}%
) is formally analogous to the Tsallis entropy $S_{\mu}$ in
Eq.~(\ref{eq-thdeEntropy}), upon identifying $\Delta\rightarrow2(\mu- 1)$.
However, this resemblance is purely mathematical as the underlying physical
origins and interpretations of the two entropies are fundamentally
different.

Barrow originally introduced $\Delta$ via a simple fractal-inspired
construction which naturally restricts it to the interval $0\leq\Delta\leq1$.
However, more general arguments indicate that negative values of the parameter
may also be allowed, thus extending the admissible range to $\Delta\in(-1,1]$;
see Refs.~\cite{refbhdedn1,refbhdedn2,refbhdedn3,refbhdedn5} for a
comprehensive discussion. Recently, the cosmological framework based on Barrow
entropy has been further generalized by allowing the deformation parameter
$\Delta$ to evolve dynamically with the energy scale, instead of remaining
constant as assumed in the original Barrow formulation
\cite{refbhdedv1,refbhdedv2,new0a}. This extension provides a broader
perspective on entropic cosmology and enriches its theoretical structure.

From the nature of the master equation (\ref{eq-OdediffeqBhde}) we
see that phenomenologically, that a bigger range for the definition of
$\Delta$ is allowed. Therefore, in this work, we consider $\Delta\in\left(
-2,2\right)  .$ The limit $\Delta\rightarrow2$, from (\ref{eq-Odbhde001})
recovers the cosmological constant limit. Therefore, the Barrow exponent
interpolates between a strongly evolving holographic component and an exact
cosmological constant. 

On the other hand, Tsallis entropy is part of nonextensive statistical
mechanics, which generalizes the Boltzmann-Gibbs formalism to systems with
long-range interactions, memory effects, or fractal structures, such as
gravitational systems. These features suggest that nonextensive behavior may
also be relevant at cosmological scales, motivating its use in the
thermodynamic description of the Universe. In such cases, the additivity
underlying Boltzmann-Gibbs entropy is generally lost. Tsallis introduced a
generalized entropy to describe these nonadditive systems. Nonetheless, the
standard Legendre structure of thermodynamics is preserved when expectation
values and internal energy are defined consistently \cite{reftebook}.

Finally, as discussed above, Tsallis holographic dark energy can be recovered
from the expressions of Barrow holographic dark energy through the
identification $\Delta\rightarrow2(\mu-1)$, or equivalently, via the
transformation $\mu\rightarrow1+\frac{\Delta}{2}$.


\section{Observational Constraints from DESI DR2}

\label{sec4}

We test our model with the late-time observational data, and specifically we
combine data given by the Cosmic Chronometers (CC), Supernova data (SNIa) and
Baryon Acoustic Oscillations (BAO). For the CC data, we consider 31 direct
measurements of the Hubble parameter $H\left(  z\right)  $ \cite{cc1} within
the redshifts range $0.09\leq z\leq1.965.~$From the difference $\frac{dz}{dt}$
between the age of galaxies at neighboring redshifts, the Hubble function
$H\left(  a\right)  =\frac{d\ln a}{dt}$ is obtained from the expression
$H\left(  z\right)  =-\frac{1}{\left(  1+z\right)  }\frac{dz}{dt}$, recall
that $\frac{1}{a}=1+z$. The corresponding likelihood is defined by using the
full covariance matrix\footnote{https://gitlab.com/mmoresco/CCcovariance} as
described in \cite{Moresco:2020fbm}. Furthermore, for the SNIa data, we
consider three different compilations, the PantheonPlus (PP)
\cite{Brout:2022vxf}, Union3.0 (U3) \cite{rubin2023union} and DES-Dovekie
(DESD) \cite{DES:2025sig}. These catalogues provide measurements of the
distance modulus $\mu^{obs}$ as a function of redshift, for the PP and U3
catalogues the redshifts are within the range $10^{-3}<z<2.27,\,\ $DESD
catalogue $z<1.13$. Finally we consider the seven redshift bins for the BAO as
provided by the Dark Energy Spectroscopic Instrument Data Release 2 (DESI DR2)
\cite{DESI:2025zpo,DESI:2025zgx,DESI:2025fii}. The bins provide seven
measurements of the comoving angular distance ratio,~the volume-averaged
distance ratio and the Hubble distance ratio, all of them normalized by the
sound horizon at the baryon drag epoch $r_{drag}$.

\subsection{Methodology \& Priors}

For the Bayesian parameter estimation of the cosmological
parameters\footnote{We have selected$~\frac{{C}}{3M_{p}^{2}}=1$.} $\left\{
H_{0},\Omega_{m0},r_{drag},\Delta,w_{m}\right\}  $, we employ
COBAYA\footnote{https://cobaya.readthedocs.io/} \cite{cob1,cob2} and a
Runge-Kutta solver to compute the cosmological evolution numerically. The
PolyChord nested sampling algorithm \cite{poly1,poly2} is used, which computes
the posterior distributions and the Bayesian evidence, necessary for the
statistical comparison of our model with the $\Lambda$CDM. Indeed, for the
statistical comparison of our model with the $\Lambda$CDM we make use of the
Akaike Information Criterion (AIC) \cite{AIC} and the Bayesian evidence
\cite{AIC2}\thinspace. For the first one, we use the difference $\Delta AIC$
of the $AIC~$parameter defined by the relation%
\[
\Delta AIC=\chi_{\min}^{2~BHDE}-\chi_{\min}^{2~\Lambda CDM}+2\left(
\mathcal{N}^{BHDE}-\mathcal{N}^{\Lambda CDM}\right)  ,
\]
in which $\mathcal{N}$ denote the degrees of freedom for each model.
Therefore, $\mathcal{N}^{BHDE}-\mathcal{N}^{\Lambda CDM}=2.$ Thus, from the
Akaike's scale, if $\left\vert \Delta AIC\right\vert <2$, the two models are
statistically equivalent, for $\left\vert \Delta AIC\right\vert <6$, there is
weak evidence in favor of the model with smaller $AIC$ and for $\left\vert
\Delta AIC\right\vert >6$, the evidence is strong. Nevertheless, the Akaike
Information Criterion is based on the goodness-of-fit value, i.e. $\chi_{\min
}^{2}$ and cannot capture the full statistical complexity difference between
models, because it does not account for the volume of the parameter space or
the prior information. On the other hand, the Bayesian evidence, integrates
the likelihood over the entire parameter space and provides a more robust
framework for model comparison, naturally balancing goodness-of-fit against
model complexity through the Occam penalty. Therefore, according to the
Jeffrey's scale \cite{AIC2} for the difference $\Delta\left(  \ln Z\right)
=\ln\frac{Z_{BHDE}}{Z_{\Lambda CDM}}$, we conclude, that if $\left\vert
\Delta\left(  \ln Z\right)  \right\vert <1$, the two models are comparable and
the data do not have a preferred model. However, when $\left\vert
\Delta\left(  \ln Z\right)  \right\vert >1$, there is and evidence in favor of
the model with the higher Bayesian evidence value, if \ $\left\vert
\Delta\left(  \ln Z\right)  \right\vert <2.5$, the evidence is weak, if
$2.5<\left\vert \Delta\left(  \ln Z\right)  \right\vert <5~$the evidence is
moderate, while for $\left\vert \Delta\left(  \ln Z\right)  \right\vert >5$,
there is very strong evidence that the given dataset supports the model with
the higher value for the Bayesian evidence.

\bigskip%
\begin{table}[tbp] \centering
\caption{Priors of the Free Parameters.}%
\begin{tabular}
[c]{ccc}\hline\hline
\textbf{Priors} & \textbf{BHDE} & $\Lambda$\textbf{CDM}\\\hline
$\mathbf{H}_{0}$ & $\left[  60,80\right]  $ & $\left[  60,80\right]  $\\
$\mathbf{\Omega}_{m0}$ & $\left[  0.01,0.5\right]  $ & $\left[
0.01,0.5\right]  $\\
$\mathbf{r}_{drag}$ & $\left[  120,170\right]  $ & $\left[  120,170\right]
$\\
$\mathbf{\Delta}$ & $\left(  -2,2\right)  $ & $-$\\
$\mathbf{w}_{m}$ & $\left[  -0.1,0.1\right]  $ & $-$\\\hline\hline
\end{tabular}
\label{prior}%
\end{table}%

\subsection{Results}

\begin{figure}[t]
\centering\includegraphics[width=0.5\textwidth]{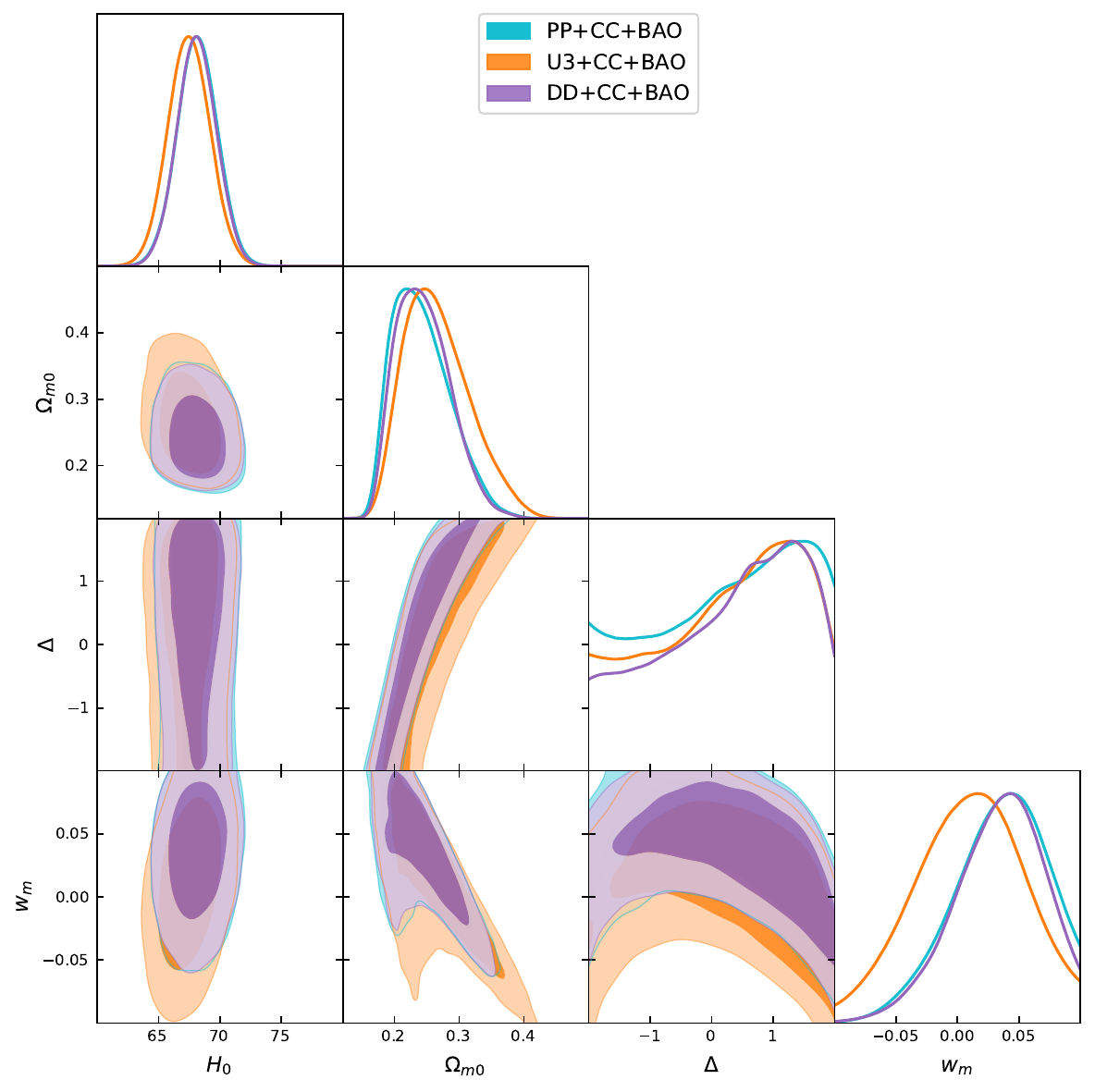}\caption{Marginalized
posterior distributions for the cosmological parameters $H_{0},~\Omega
_{m0},~\Delta$ and $w_{m}$ obtained from the Bayesian analysis. The contours
correspond to the 68\% and 95\% CI.}%
\label{fig1}%
\end{figure}

We consider the following combined datasets PP+CC+BAO,~U3+CC+BAO and DD+CC+BAO
and priors for the parameters $\left\{  H_{0},\Omega_{m0},r_{drag}%
,\Delta,w_{m}\right\}  $ as given in Table \ref{prior}. Recall that for the
$\Omega_{b0}=\frac{\omega_{b}}{h^{2}},$ we consider the value obtained from
the Planck 2018 collaboration \cite{refplanck}. The analysis of the numerical
chains was performed by using the GetDist library \cite{get}, and the obtained
median parameters of the posterior space, and the 68\% confidence
intervals\ (CI) are presented in Table \ref{tab2}, while the marginalized
posterior distributions for the cosmological parameters for the 68\% and 95\%
CI are given in Fig. \ref{fig1}.%

\begin{table*}[tbp] \centering
\caption{Numerical outcomesfor the BHDE model.}%
\begin{tabular}
[c]{cccc}\hline\hline
\textbf{Dataset} & \textbf{PP+CC+BAO} & \textbf{U3+CC+BAO} &
\textbf{DD+CC+BAO}\\\hline
$\mathbf{H}_{0}~\left[  \mathrm{km/s/Mpc}\right]  $ & $68.2_{-1.6}^{+1.6}$ &
$67.5_{-1.7}^{+1.7}$ & $68.1_{-1.6}^{+1.6}$\\
$\mathbf{\Omega}_{m0}$ & $0.243_{-0.056}^{+0.030}$ & $0.266_{-0.062}^{+0.038}$
& $0.245_{-0.050}^{+0.033}$\\
$\mathbf{r}_{drag}~\left[  \mathrm{Mpc}\right]  $ & $146.8_{-3.3}^{+3.3}$ &
$146.8_{-3.3}^{+3.3}$ & $146.8_{-3.3}^{+3.3}$\\
$\mathbf{\Delta}$ & $-$ & $0.26_{-0.58}^{+1.7}$ & $0.31_{-0.56}^{+1.6}$\\
$\mathbf{w}_{m}$ & $0.033_{-0.030}^{+0.045}$ & $0.009_{-0.041}^{+0.047}$ &
$0.033_{-0.029}^{+0.042}$\\
$\Delta\mathbf{\chi}_{\min}^{2}$ & $-1.77$ & $-3.25$ & $-2.62$\\
$\Delta\mathbf{AIC}$ & $+2.23$ & $+0.75$ & $+1.38$\\
$\Delta\ln\mathbf{Z}$ & $-0.63$ & $+0.47$ & $-0.13$\\\hline\hline
\end{tabular}
\label{tab2}%
\end{table*}%
%

\begin{table*}[tbp] \centering
\caption{Comparison of the statistical parameters for the BHDE model with respect to the $\Lambda$CDM.}%
\begin{tabular}
[c]{ccccc}\hline\hline
\textbf{Dataset} & $\Delta\mathbf{AIC}$ & \textbf{Akaike's Scale} & $\Delta
\ln\mathbf{Z}$ & \textbf{Jeffrey's Scale}\\\hline
\textbf{PP+CC+BAO} & $+2.23$ & Weak evidence for $\Lambda$CDM & $-0.63$ &
Inconclusive\\
\textbf{U3+CC+BAO} & $+0.75$ & Inconclusive & $+0.47$ & Inconclusive\\
\textbf{DD+CC+BAO} & $+1.38$ & Inconclusive & $-0.13$ &
Inconclusive\\\hline\hline
\end{tabular}
\label{tab3}%
\end{table*}%

The numerical results of Table~\ref{tab2} suggests that the constraints on the
$H_{0}$ are consistent and independent from the choice of the SNIa catalogue.
The three datasets suggests $H_{0}=68.2_{-1.6}^{+1.6}$, $67.5_{-1.7}^{+1.7}$
and $68.1_{-1.6}^{+1.6}~\mathrm{km/s/Mpc}$ for the combined datasets
PP+CC+BAO, U3+CC+BAO and DD+CC+BAO respectively. The sound horizon at the drag
epoch, $r_{drag}=146.8_{-3.3}^{+3.3}~\mathrm{Mpc}$, is recover to have the
same value for all the different combinations. On the other hand, the density
parameter for the dark matter component is constraint to be within the range
$\Omega_{m0}\simeq0.24-0.27$, with central values below the value obtained for
the $\Lambda$CDM. This difference, is due to the new parameter $w_{m}$, since
a positive $w_{m}$ makes the dark matter component dilute faster than $\left(
1+z\right)  ^{3}$, so that a smaller $\Omega_{m0}$ is required in order to
reproduce the same expansion history at intermediate redshifts. \ Parameter
$w_{m}$ is constraints as $w_{m}=0.033_{-0.030}^{+0.045}$, $0.009_{-0.041}%
^{+0.047}$ and $0.033_{-0.029}^{+0.042}$, that is, the pressureless limit
$w_{m}=0$ is recovered within the $1\sigma~$for the U3+CC+BAO data and within
the $2\sigma$ for the other catalogues, thus, there is not a strong evidence
for a nonzero pressure term in the dark matter. Finally, parameter $\Delta$ of
the BHDE model is the least constrained of the set. For U3+CC+BAO and
DD+CC+BAO we obtain $\Delta=0.26_{-0.58}^{+1.7}$ and $\Delta=0.31_{-0.56}%
^{+1.6}$, that is, while for PP+CC+BAO the posterior remains flat and no
meaningful interval can be quoted. It should be emphasized that in the two
former cases the upper end of the interval $\Delta\simeq1.96$ is very close to
the upper limit $\Delta\rightarrow2$, where the cosmological constant is
recovered. On the other hand, the data suggest $\Delta>-0.3$ for the $68\%$~CI.

The background data employed here are therefore able to exclude only the
region of large deviations, and the determination of $\Delta$ requires
observables that are sensitive to the perturbations, or to higher redshifts
such that to break the degeneracy of the parameters $\Delta-w_{m}$ as can be
seen in Fig. \ref{fig1}.

\begin{figure*}[t]
\centering\includegraphics[width=1\textwidth]{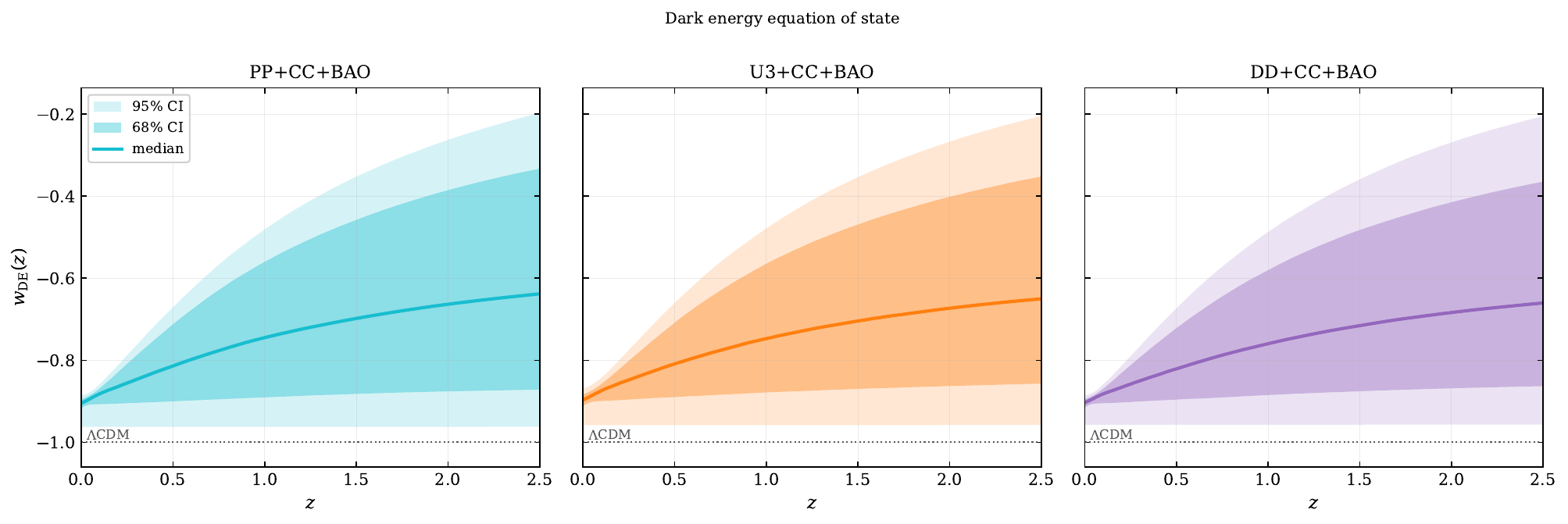}\caption{Qualitative
evolution of the dark energy equation of state parameter $w_{d}\left(
z\right)  $ and the $68\%$ and $95\%$ CI. }%
\label{fig2}%
\end{figure*}

Finally, concerning the statistical comparison with the $\Lambda$CDM model,
the BHDE model provides a better-fit in all three cases, with $\Delta
\chi_{\min}^{2}=-1.77$, $-3.25$ and $-2.62$,~as given in\ Table \ref{tab3}.
Nevertheless, such an improvement is what one expects from the introduction of
two additional number of free parameters. Hence from the Akaike Information
Criterion we calculate $\Delta\mathrm{AIC}=+2.23$, $+0.75$ and $+1.38$, so
that $\Lambda$CDM to be weakly preferred by the PP data, while for the other
two datasets the two models are statistically equivalent. \ Finally, for the
difference of the Bayesian evidence with the $\Lambda$CDM is calculated
$\Delta\ln Z=-0.63$, $+0.47$ and $-0.13$, which means that the two models fit
the data in a similar model and are statistically equivalent.

In Fig. \ref{fig2} we present the qualitative evolution for the dark energy
equation of state parameter $w_{d}\left(  z\right)  $ as follows from the
constraint values of Table \ref{tab2}. We observe that for this model, the
dark energy equation of state does not cross the phantom divide line, which is
in agreement with the result obtained before for $w_{m}=0$.

\begin{figure*}[t]
\centering\includegraphics[width=1\textwidth]{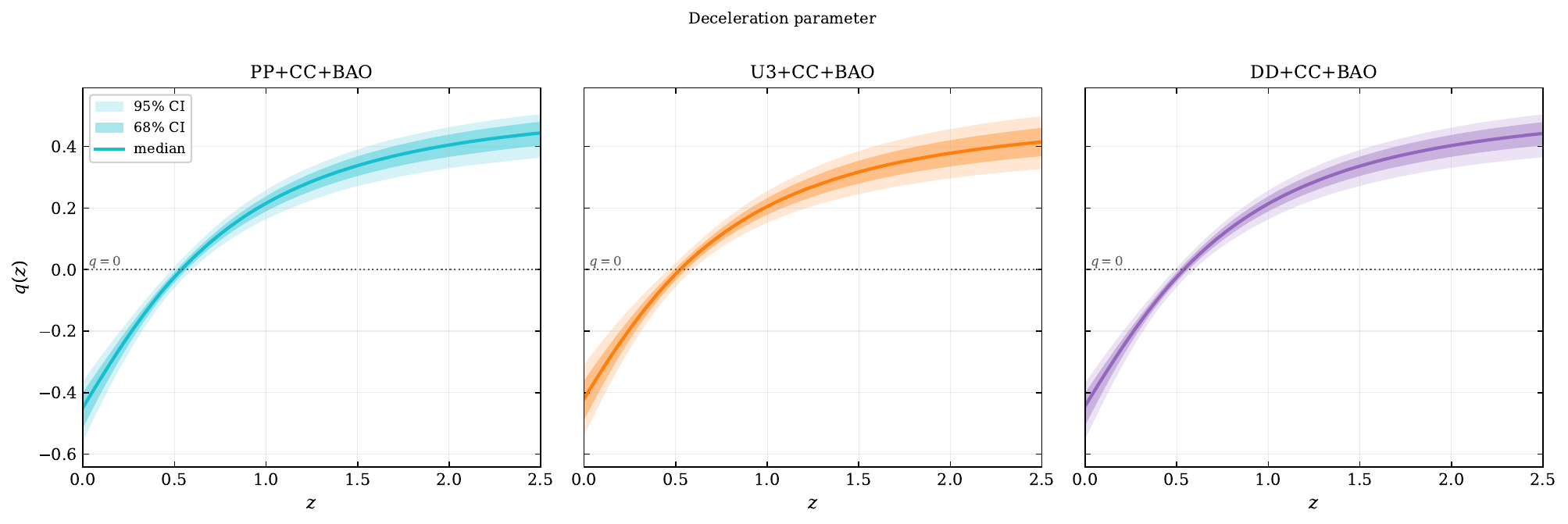}\caption{Qualitative
evolution of the deceleration parameter $q\left(  z\right)  $ and the $68\%$
and $95\%$ CI. }%
\label{fig4}%
\end{figure*}

\begin{figure*}[t]
\centering\includegraphics[width=1\textwidth]{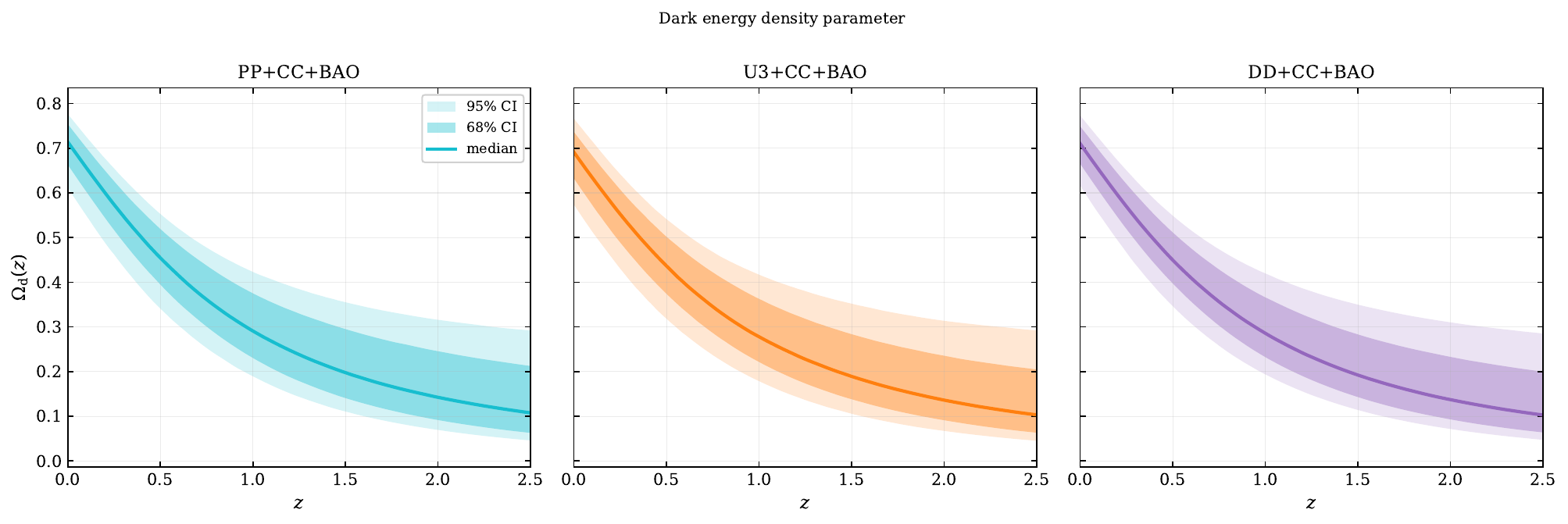}\caption{Qualitative
evolution of thedark energy density $\Omega_{d}\left(  z\right)  $ and the
$68\%$ and $95\%$ CI. }%
\label{fig5}%
\end{figure*}

Finally, the obtained cosmological background functions, such as the deceleration parameter $q\left(  z\right)
$, and the dark energy density $\Omega_{d}\left(  z\right)  $ are given in
Figs. \ref{fig4} and \ref{fig5} respectively.

\section{Conclusions}

\label{sec5}

In this work we have considered the BHDE scenario in a spatially flat universe
whose dark sector is dynamical on both sides.\ Specifically, the dark energy
density is fixed by a Barrow-deformed entropy--area relation with the future
event horizon as infrared cutoff, while the dark matter component is endowed
with a constant, non-zero equation-of-state parameter, that is, the dark
energy has a nonzero pressure contribution to the cosmic fluid. 

For this model, we have derived the master equation which describe the
dynamics for the background evolution. The equation was solved numerically,
and we employed the COBAYA Bayesian interference, to confront the cosmological
model with late-time observational data. In particular we employed background
data, and specifically the\ BAO measurements given by the second release of
DESI, the CC and SNIa data. \ In this work, we considered the Barrow exponent
$\Delta$ to have the limit $\Delta\rightarrow2$, where the case of the
cosmological constant is recovered. \ Thus, the exponent $\Delta$ can describe
a rapidly evolving holographic dark energy for small values of $\Delta$, or
the cosmological constant when $\Delta\rightarrow2$. The cosmological data,
provided weak fits on the parameter $\Delta\,$, suggesting a degeneracy with
the dark matter equation of state parameter $w_{m}$. Thus, the cosmological
constant limit  $\Delta\rightarrow2$, to be within the $1\sigma$ and $2\sigma$
regimes, as presented in Fig. \ref{fig1}. This explains the obtained values
for the comparison of the Bayesian evidence, $\Delta\ln Z=-0.63$, $+0.47$ and
$-0.13,$ between the BHDE and the Bayesian evidence. 

Our analysis is restricted to the background evolution, and does not account
on the effects of the nonzero $w_{m}~$on the sound speed for the perturbation
equations. Furthermore, the degeneracy between parameters $\Delta-w_{m}$ can
not break from these background data, and cosmological perturbations should be
introduced. In a future work we plan to work within this direction and address
the effects of the nonzero $w_{m}$, as also to explore the deviation of the
BHDE from the cosmological constant limit. 

\begin{acknowledgments}
GL thanks the support of Vicerrector\'{\i}a de Investigaci\'{o}n y Desarrollo
Tecnol\'{o}gico (VRIDT) of Universidad Cat\'{o}lica del Norte (UCN) through
Resoluci\'{o}n VRIDT No. 096/2022, Resoluci\'{o}n VRIDT No. 200/2025 and
Resoluci\'{o}n VRIDT No. 021/2026. Part of this study was supported by
FONDECYT Grant 1240514.
\end{acknowledgments}



\begin{thebibliography}{99}                                                                                               %


\bibitem {refplanck}N. Aghanim \textit{et al.} [Planck Collaboration],
\textquotedblleft Planck 2018 results VI. Cosmological
parameters,\textquotedblright\ Astron. Astrophys. 641, A6 (2020) [Erratum:
Astron. Astrophys. 652, C4 (2021)]

\bibitem {refdm01}R. H. Wechsler and J. L. Tinker, ``The Connection between
Galaxies and their Dark Matter Halos,'' Ann. Rev. Astron. Astrophys.
\textbf{56}, 435 (2018)

\bibitem {acc1}A.~G.~Riess \textit{et al.} [Supernova Search Team],
``Observational evidence from supernovae for an accelerating universe and a
cosmological constant,'' Astron. J. \textbf{116}, 1009 (1998)

\bibitem {acc2}S.~Perlmutter \textit{et al.} [Supernova Cosmology Project],
``Measurements of $\Omega$ and $\Lambda$ from 42 High Redshift Supernovae,''
Astrophys. J. \textbf{517}, 565 (1999)

\bibitem {Perivolaropoulos1}L. Perivolaropoulos and F. Skara, ``Challenges for
$\Lambda$CDM: An update," New Astron. Rev. \textbf{95}, 101659 (2022)


\bibitem {refhp1}G.~'t Hooft, ``Dimensional reduction in quantum gravity,''
Conf. Proc. C \textbf{930308}, 284 (1993)

\bibitem {refhp2}L.~Susskind, ``The World as a hologram,'' J. Math. Phys.
\textbf{36}, 6377 (1995)

\bibitem {refhp3}A.~G.~Cohen, D.~B.~Kaplan and A.~E.~Nelson, ``Effective field
theory, black holes, and the cosmological constant,'' Phys. Rev. Lett.
\textbf{82}, 4971 (1999)

\bibitem {refshde1}M.~Li, ``A Model of holographic dark energy,'' Phys. Lett.
B \textbf{603}, 1 (2004)


\bibitem {refbarrowen}J.~D.~Barrow, ``The Area of a Rough Black Hole,'' Phys.
Lett. B \textbf{808}, 135643 (2020)

\bibitem {refte1}C.~Tsallis and L.~J.~L.~Cirto, \textquotedblleft Black hole
thermodynamical entropy,\textquotedblright\ Eur. Phys. J. C \textbf{73}, 2487 (2013)

\bibitem {refte2}C. Tsallis and L. J. L. Cirto, ``Black hole thermodynamical
entropy,'' Eur. Phys. J. C \textbf{73}, 2487 (2013)

\bibitem {reftebook}C. Tsallis, \textit{Introduction to Non-Extensive
Statistical Mechanics: Approaching a Complex World} (Springer, Berlin, 2009).

\bibitem {refve1}S.~Viaggiu, ``Bekenstein-Hawking entropy in expanding
universes from black hole theorems,'' Mod. Phys. Lett. A \textbf{29}, 1450091 (2014)

\bibitem {refve2}S.~Viaggiu, ``First law of thermodynamics for dynamical
apparent horizons and the entropy of Friedmann universes,'' Gen. Rel. Grav.
\textbf{47}, 86 (2015)


\bibitem {refbhde}E.~N.~Saridakis, ``Barrow holographic dark energy,'' Phys.
Rev. D \textbf{102}, 123525 (2020)

\bibitem {refbhde2}F.~K.~Anagnostopoulos, S.~Basilakos, and E.~N.~Saridakis,
``Observational constraints on Barrow holographic dark energy,'' Eur. Phys. J.
C \textbf{80}, 826 (2020)

\bibitem {refbhde3}A.~A.~Mamon, A.~Paliathanasis, and S.~Saha, ``Dynamics of
an Interacting Barrow Holographic Dark Energy Model and its Thermodynamic
Implications,'' Eur. Phys. J. Plus \textbf{136}, 134 (2021)

\bibitem {refbhde4}P.~Adhikary, S.~Das, S.~Basilakos, and E.~N.~Saridakis,
``Barrow holographic dark energy in a nonflat universe,'' Phys. Rev. D
\textbf{104}, 123519 (2021)

\bibitem {refbhde5}A.~Al Mamon, A.~K.~Mishra and U.~K.~Sharma, ``Barrow
Holographic dark energy in fractal cosmology,'' Int. J. Geom. Meth. Mod. Phys.
\textbf{19}, 2250231 (2022)

\bibitem {refbhde6}A.~Al Mamon, U.~K.~Sharma, M.~Kumar, and A.~K.~Mishra,
\textquotedblleft Cosmic consequences of Barrow holographic dark energy with
Granda-Oliveros cut-off in fractal cosmology,\textquotedblright\ Gen. Rel.
Grav. \textbf{55}, 74 (2023)

\bibitem {refbhde7}G.~G.~Luciano, A.~Paliathanasis, and E.~N.~Saridakis,
``Barrow and Tsallis entropies after the DESI DR2 BAO data,'' 
\textbf{09}, 013 (2025)

\bibitem {refbhde8}G.~G.~Luciano, A.~Paliathanasis, and E.~N.~Saridakis,
``Constraints on Barrow and Tsallis holographic dark energy from DESI DR2 BAO
data,'' JHEAp \textbf{49}, 100427 (2026)

\bibitem {refbhdedv1}S.~Di Gennaro and Y.~C.~Ong, ``Sign Switching Dark Energy
from a Running Barrow Entropy,'' Universe \textbf{8}, 541 (2022)

\bibitem {refbhdedv2}S.~Basilakos, A.~Lymperis, M.~Petronikolou, and
E.~N.~Saridakis, ``Barrow holographic dark energy with varying exponent,''
Nucl. Phys. B \textbf{1015}, 116904 (2025)

\bibitem {refthde1}M.~Tavayef, A.~Sheykhi, K.~Bamba, and H.~Moradpour,
``Tsallis Holographic Dark Energy,'' Phys. Lett. B \textbf{781}, 195 (2018)

\bibitem {refthde2}M.~A.~Zadeh, A.~Sheykhi, H.~Moradpour, and K.~Bamba, ``Note
on Tsallis holographic dark energy,'' Eur. Phys. J. C \textbf{78}, 940 (2018)

\bibitem {refthde3}S.~Ghaffari, H.~Moradpour, V.~B.~Bezerra, J.~P.~Morais
Gra{\c{c}}a, and I.~P.~Lobo, ``Tsallis holographic dark energy in the brane
cosmology,'' Phys. Dark Univ. \textbf{23}, 100246 (2019)

\bibitem {refthde4}M.~Abdollahi Zadeh, A.~Sheykhi, and H.~Moradpour, ``Thermal
stability of Tsallis holographic dark energy in nonflat universe,'' Gen. Rel.
Grav. \textbf{51}, 1 (2019)

\bibitem {refthde5}A.~Al~Mamon, A.~H.~Ziaie, and K.~Bamba, ``A generalized
interacting Tsallis holographic dark energy model and its thermodynamic
implications,'' Eur. Phys. J. C \textbf{80}, 974 (2020)

\bibitem {refthde6}U.~K.~Sharma, N.~M.~Ali, A.~Al Mamon and Pankaj,
``Interacting New Tsallis holographic dark energy,'' Chin. J. Phys.
\textbf{89}, 657 (2024)

\bibitem {refthde7}A.~Al Mamon,``Study of Tsallis holographic dark energy
model in the framework of Fractal cosmology,'' Mod. Phys. Lett. A \textbf{35},
2050251 (2020)

\bibitem {refvhde1}S.~Saha, S.~Saha, and N.~Mahata, ``The peculiar case of the
Viaggiu holographic dark energy,'' Nucl. Phys. B \textbf{1025}, 117368 (2026)

\bibitem {refvhde2}A.~K.~Halder, A.~Paliathanasis, S.~Viaggiu, A.~A.~Mamon,
and S.~Saha, ``Viaggiu holographic dark energy in light of DESI DR2,'' Phys.
Lett. B \textbf{876}, 140442 (2026)

\bibitem {refvhde3}A.~Jhunjhunwala and S.~Maity, ``Reconstructing $f(R)$
gravity from Viaggiu Holographic Dark Energy under Hubble, Event Horizon and
Granda Oliveros cutoffs,'' arXiv:2607.03308 [gr-qc]

\bibitem {refshde2}S.~Wang, Y.~Wang, and M.~Li, ``Holographic Dark Energy,''
Phys. Rept. \textbf{696}, 1 (2017)

\bibitem {Nojiri1}S. Nojiri, S. D. Odintsov, and T. Paul, ``Horizon entropy
consistent with the FLRW equations for general modified theories of gravity
and for all equations of state of the matter field," Phys. Rev. D
\textbf{109}, 043532 (2024)

\bibitem {Nojiri2}S. Nojiri, S. D. Odintsov, and T. Paul, ``Different aspects
of entropic cosmology," Universe \textbf{10}, 352 (2024)

\bibitem {Luciano1}G. G. Luciano and A. Paliathanasis, ``Modified cosmology
through generalized mass-to-horizon entropy: Observational constraints from
DESI DR2 BAO data," Phys. Lett. B \textbf{870}, 139954 (2025)

\bibitem {Luciano2}G.~G.~Luciano and E.~N.~Saridakis, ``New modified cosmology
from a new generalized entropy," Phys. Lett. B \textbf{879}, 140703 (2026)

\bibitem {Leizerovich}M.~Leizerovich, S.~J.~Landau, G.~G.~Luciano,
A.~Papatriantafyllou and E.~N.~Saridakis, ``Observational constraints on
Luciano-Saridakis entropic cosmology," Phys. Dark Univ. \textbf{52}, 102333 (2026)


\bibitem {desidr22dde}D.~Wang and D.~Mota, ``Did DESI DR2 truly reveal
dynamical dark energy?,'' Eur. Phys. J. C \textbf{85}, 1356 (2025)

\bibitem {ddmmoti1}F. Zwicky, ``Die Rotverschiebung von extragalaktischen
Nebeln,'' Helv. Phys. Acta \textbf{6}, 110 (1933)

\bibitem {ddmmoti2}W. Hu, ``Structure formation with generalized dark
matter,'' Astrophys. J. \textbf{506}, 485-494 (1998).

\bibitem {ddm1}C.~M.~Muller, ``Cosmological bounds on the equation of state of
dark matter,'' Phys. Rev. D \textbf{71}, 047302 (2005)



\bibitem {ddm2}S. Kumar and L. Xu, ``Observational constraints on variable
equation of state parameters of dark matter and dark energy after Planck,''
Phys. Lett. B \textbf{737}, 244 (2014)

\bibitem {ddm3}S. Kumar, R. C. Nunes, and S. K. Yadav, ``Testing the warmness
of dark matter,'' Mon. Not. Roy. Astron. Soc. \textbf{490}, 1406 (2019)

\bibitem {ddm4}M.~Kopp, C.~Skordis, and D.~B.~Thomas, ``Extensive
investigation of the generalized dark matter model,'' Phys. Rev. D
\textbf{94}, 043512 (2016)

\bibitem {ddm5}S.~Gariazzo, M.~Escudero, R.~Diamanti, and O.~Mena,
``Cosmological searches for a noncold dark matter component,'' Phys. Rev. D
\textbf{96}, 043501 (2017)

\bibitem {ddm6}R.~Murgia, A.~Merle, M.~Viel, M.~Totzauer, and A.~Schneider,
````Non-cold'' dark matter at small scales: a general approach,'' J. Cosmol. Astropart. Phys. \textbf{11}, 046 (2017)

\bibitem {ddm7}R.~Murgia, V.~Ir{\v{s}}i{\v{c}} and M.~Viel, ``Novel
constraints on noncold, nonthermal dark matter from Lyman- {$\alpha$} forest
data,'' Phys. Rev. D \textbf{98}, 083540 (2018)

\bibitem {ddm8}A.~Schneider, ``Constraining noncold dark matter models with
the global 21-cm signal,'' Phys. Rev. D \textbf{98}, 063021 (2018)

\bibitem {ddm9}M.~Kopp, C.~Skordis, D.~B.~Thomas and S.~Ili{\'c}, ``Dark
Matter Equation of State through Cosmic History,'' Phys. Rev. Lett.
\textbf{120}, 221102 (2018)

\bibitem {ddm10}S.~N{\'a}jera and R.~A.~Sussman, ``Non-comoving cold dark
matter in a $\Lambda$CDM background,'' Eur. Phys. J. C \textbf{81}, 374 (2021)

\bibitem {ddm11}S.~Ili{\'c}, M.~Kopp, C.~Skordis and D.~B.~Thomas,``Dark
matter properties through cosmic history,'' Phys. Rev. D \textbf{104}, 043520 (2021)

\bibitem {ddm12}Y.~H.~Yao, J.~C.~Wang, and X.~H.~Meng, ``Observational
constraints on noncold dark matter and phenomenological emergent dark
energy,'' Phys. Rev. D \textbf{109}, 063502 (2024)

\bibitem {ddm13}Y.~H.~Yao and J.~Q.~Liu,``Hints of noncold dark matter?
Observational constraints on barotropic dark matter with a constant equation
of state parameter,'' Phys. Dark Univ. \textbf{49}, 102052 (2025)

\bibitem {ddm14}T.~N.~Li, Y.~M.~Zhang, Y.~H.~Yao, G.~H.~Du, P.~J.~Wu,
J.~F.~Zhang, and X.~Zhang, ``Revisiting the phenomenologically emergent dark
energy model: is non-zero equation of state of dark matter favored by DESI
DR2?,'' J. Cosmol. Astropart. Phys. \textbf{12}, 048 (2025)

\bibitem {ddm15}Y.~Carloni and O.~Luongo,\textquotedblleft A barotropic
alternative to Early Dark Energy for alleviating the $H_{0}$
tension,\textquotedblright\ arXiv:2604.18053 [astro-ph.CO]

\bibitem {ddm16}M. Abedin, L. A. Escamilla, S. Pan, E. Di Valentino and W.
Yang, When dark matter heats up: A model-independent search for noncold
behavior, Phys. Rev. D \textbf{112}, 123537 (2025) 

\bibitem {ddm17}W. Yang, S. Pan, E. Di Valentino, O. Mena and D. F. Mota,
"Probing the cold nature of dark matter", Phys.\ Rev. D \textbf{111}, 103509 (2025)

\bibitem {ddmpara}D.~Wang, ``Evidence for Dynamical Dark Matter,''
arXiv:2504.21481 [astro-ph.CO]

\bibitem {refliddm}T.~N.~Li, P.~J.~Wu, G.~H.~Du, Y.~H.~Yao, J.~F.~Zhang, and
X.~Zhang, ``Exploring non-cold dark matter in the scenario of dynamical dark
energy with DESI DR2 data,'' Phys. Dark Univ. \textbf{50}, 102068 (2025)


\bibitem {de1}E.~J.~Copeland, M.~Sami, and S.~Tsujikawa, ``Dynamics of dark
energy,'' Int. J. Mod. Phys. D \textbf{15}, 1753 (2006)


\bibitem {refbhe1}J.~D.~Bekenstein, ``Black holes and the second law,'' Lett.
Nuovo Cim. \textbf{4} (1972), 737-740

\bibitem {refbhe2}J.~D.~Bekenstein, ``Black holes and entropy,'' Phys. Rev. D
\textbf{7}, 2333 (1973)

\bibitem {refbhdedn1}E.~Dagotto, A.~Kocic, and J.~B.~Kogut, ``Collapse of the
wave function, anomalous dimensions and continuum limits in model scalar field
theories,'' Phys. Lett. B \textbf{237}, 268 (1990)

\bibitem {refbhdedn2}P. Xu, B. Yu, ``Developing a new form of permeability and
Kozeny-Carman constant for homogeneous porous media by means of fractal
geometry,'' Advances in Water Resources \textbf{31}, 74 (2008)

\bibitem {refbhdedn3}H. H. P. Tang, J. Z. Wang, J. L. Zhu, and Q. B. Ao,
``Fractal dimension of pore-structure of porous metal materials made by
stainless steel powder,'' Powder Tech. \textbf{217}, 383 (2012).



\bibitem {refbhdedn5}P.~Jizba, G.~Lambiase, G.~G.~Luciano, and
L.~Mastrototaro, \textquotedblleft Imprints of Barrow{-}Tsallis cosmology in
primordial gravitational waves,\textquotedblright\ Eur. Phys. J. C
\textbf{84}, 1076 (2024)



\bibitem {new0a}G. G. Luciano, O. Luongo and M. Muccino, "Cosmological
consequences of scale-dependent Barrow-Tsallis entropy", arXiv:2607.26105 [gr-qc] (2026)

\bibitem {cc1}S. Vagnozzi, A. Loeb, and M. Moresco, ``Eppur \'e piatto? The cosmic chronometer take on spatial curvature and cosmic concordance," Astrophys. J. \textbf{908}, 84 (2021)

\bibitem {Moresco:2020fbm}M.~Moresco, R.~Jimenez, L.~Verde, A.~Cimatti, and
L.~Pozzetti, ``Setting the stage for cosmic chronometers. II. Impact of stellar population synthesis models systematics and full covariance matrix,” Astrophys. J. \textbf{898}, 82 (2020)

\bibitem {Brout:2022vxf}D. Brout \textit{et al.}, ``The Pantheon+ analysis: cosmological constraints,” Astrophys. J. \textbf{938}, 110 (2022)

\bibitem {rubin2023union}D. Rubin \textit{et al.}, ``Union through UNITY: Cosmology with 2,000 SNe using a unified Bayesian framework,” Astrophys. J. \textbf{986}, 231 (2025)

\bibitem {DES:2025sig}B. Popovic \textit{et al.} (DES Collaboration), ``The Dark Energy Survey Supernova Program: A Reanalysis Of Cosmology Results
And Evidence For Evolving Dark Energy With An Updated Type Ia Supernova Calibration,” Mon. Not. R. Astron. Soc. \textbf{548}, 1 (2026)

\bibitem {DESI:2025zpo}M. Abdul Karim \textit{et al.} (DESI Collaboration), ``DESI DR2 results. I. Baryon acoustic oscillations from the Lyman alpha forest,'' Phys. Rev. D \textbf{112}, 083514 (2025)

\bibitem {DESI:2025zgx}M. Abdul Karim \textit{et al.} (DESI Collaboration), ``DESI DR2 results. II. Measurements of baryon acoustic oscillations and cosmological constraints,'' Phys. Rev. D \textbf{112}, 083515 (2025)

\bibitem {DESI:2025fii}K. Lodha \textit{et al.} (DESI Collaboration), ``Extended dark energy analysis using DESI DR2 BAO measurements,'' Phys. Rev. D \textbf{112}, 083511 (2025)

\bibitem {cob1}J. Torrado and A. Lewis, ``Cobaya: Bayesian analysis in cosmology," ascl:1910.019 (2019)

\bibitem {cob2}J. Torrado and A. Lewis, ``Cobaya: Code for Bayesian Analysis of hierarchical physical models,” J. Cosmol. Astropart. Phys. \textbf{05}, 057 (2021)

\bibitem {poly1}W. J. Handley, M. P. Hobson, and A. N. Lasenby, ``PolyChord: nested sampling for cosmology,” Mon. Not. R. Astron. Soc. \textbf{450}, L61 (2015)

\bibitem {poly2}W. J. Handley, M. P. Hobson, and A. N. Lasenby, ``PolyChord: next-generation nested sampling,” Mon. Not. R. Astron. Soc. \textbf{453}, 4384 (2015)

\bibitem {AIC}H. Akaike, ``A new look at the statistical model identification," IEEE Trans. Automat. Contr. \textbf{19}, 716 (1974)

\bibitem {AIC2}H. Jeffreys, \textit{Theory of Probability,} (Oxford University
Press, Oxford, 1961)

\bibitem {get}A. Lewis, ``GetDist: a Python package for analysing Monte Carlo samples,” J. Cosmol. Astropart. Phys. \textbf{08}, 025 (2019)
\end{thebibliography}
\end{document}